\documentstyle[amssymb,epsfig,12pt]{article}

\input{tcilatex}

\begin{document}

\bigskip \bigskip 
\begin{titlepage}
\bigskip \begin{flushright}
WATPHYS-TH02/07\\
hep-th/yymmddd
\end{flushright}


\vspace{1cm}

\begin{center}
{\Large \bf {Expanding the Area of Gravitational Entropy}}\\
\end{center}
\vspace{2cm}
\begin{center}
R. B. Mann
\footnote{
EMail: mann@avatar.uwaterloo.ca} \\
Department of Physics, University of Waterloo, \\
Waterloo, Ontario N2L 3G1, CANADA\\
\vspace{1cm}
\today\\
\end{center}
\begin{abstract}
I describe how gravitational entropy is intimately connected with the concept of 
gravitational heat, expressed as the difference between the total and free energies of
a given gravitational system.  From this perspective one can compute these
thermodyanmic quantities in settings that go considerably beyond Bekenstein's 
original insight that the area of a black hole event horizon can be identified 
with thermodynamic entropy.
The settings include the outsides of cosmological horizons and spacetimes with 
NUT charge.  However the interpretation of gravitational entropy in these broader
contexts remains to be understood.
\end{abstract}
\end{titlepage}\onecolumn   

\section{Introduction}

It would not be an exaggeration to say that the two major achievements of
20th century physics -- quantum mechanics and general relativity -- have
left us with some of science's most baffling conundrums. From the role of
time to the nature of measurement to the definition of a particle, their
conceptual foundations are markedly different and are apparently at odds
with one another. \ The spectacular empirical success enjoyed by each of
these two pillars of theoretical physics has only served to deepen the
conundrums, in large part because there is very little common territory
shared by each.

It was a key achievement of Jacob Bekenstein to pioneer such common
territory. \ In proposing that the area of a black hole was akin to
thermodynamic entropy \cite{BekH}, he opened up the new subfield of
gravitational thermodynamics, in which quantum mechanics and general
relativity can simultaneously be put to the test. Although up until now this
has been possible only via gedanken experiments, a small but growing body of
researchers are seriously entertaining the possibility of creating black
holes in a laboratory setting \cite{GiddBH}. \ I shall not explore this
tantalizing yet highly speculative idea further here, but the fact that it
is being considered at all shows how far the subject has come in the three
decades since Bekenstein's original paper. \ 

Once Hawking established that black holes could radiate a thermal spectrum
of particles via quantum-field-theoretic effects \cite{HawkBH}, the
relationship between entropy and area that Bekenstein pointed out became a
pillar of gravitational thermodynamics. \ Since the entropy/area
relationship does not exist unless $\hslash $ is nonzero, it became a key
pivot point in the development of a quantum theory of gravity. Although
progress was somewhat slow at first, an enormous amount of effort has been
expended in the last decade by the theoretical physics community in
understanding the origin of black hole entropy. This research was carried
out along two main lines. \ One involved an investigation of the breadth of
situations in which the first law of gravitational thermodynamics applies,
including other dimensionalities, dilaton gravity, cosmological settings,
higher-derivative theories and string theory. \ \ \ The other concentrated
on finding a statistical-mechanical interpretation of black hole entropy by
searching for the underlying degrees of freedom in a candidate quantum
theory of gravity such as loop quantum gravity or string theory. \ Many
important results on gravitational entropy have been achieved, including its
quasilocal formulation \cite{BY1}, its Noether-charge interpretation \cite%
{Noether}, obtaining and controlling its quantum corrections \cite{BHquant},
solvability of toy quantum gravity models \cite{twodmodels}, its
relationship to pair-production of black holes \cite{BHpair,MannRoss}, and a
partial understanding of its underlying degrees of freedom in terms of
string modes \cite{BHstring}, spin-networks \cite{BHspinnet}, and boundary
diffeomorphisms \cite{BHbound}. \ 

Rather than review each of these developments here, my purpose in this paper
is to take a small step along the path Bekenstein forged by proposing that
gravitational entropy must be present whenever gravitational heat can be
defined. \ The `heat' referred to here is that given by the Gibbs-Duhem
relation, which expresses the heat for any thermodynamic system in terms of
the difference between its total energy and its Helmholtz free energy. \
This relationship was first explored in the context of black holes by
Gibbons and Hawking \cite{GibH}, who argued that the Euclidean gravitational
action is equal to the grand canonical free energy times the reciprocal of
the temperature associated with a black hole event horizon. It is natural to
more generally claim that whenever the difference between the total and free
energies is nonzero in a given gravitational setting there will be heat,
that form of energy which is useless for doing work. \ Provided that an
equilibrium temperature can be defined (at least approximately), there will
then be a gravitational entropy, given by the ratio of the gravitational
heat energy to this temperature. \ 

\bigskip

\section{Gravitational Heat}

\bigskip

Any thermodynamic system at some constant temperature $T=\beta ^{-1}$\ will
have a total internal energy $U$ that can be meaningfully partitioned into
two parts: the amount available to do work on some other system (the free
energy or Helmholtz potential $F$), and the remainder, referred to as the
heat. \ This relationship is expressed as 
\begin{equation}
U=F+\beta ^{-1}S  \label{helm1}
\end{equation}%
where $S$ is referred to as the entropy of the system. \ From this
perspective, entropy can be regarded as being defined by the ratio of the
difference between the total and free energies of the system with its
temperature. \ Alternatively, by taking differentials of eq. (\ref{helm1}),
the relation $S=\beta ^{2}\frac{\partial F}{\partial \beta }$ is easily
obtained. \ 

Other thermodynamic potentials exist of course -- the Gibbs potential, the
enthalpy, etc. However the advantage of the Helmholtz free energy is that it
admits a straightforward generalization to gravitation via the (Euclidean)
path-integral formalism \cite{GibH}. In other words, one can define
gravitational heat to be the difference between the total and free energies
of a given gravitational system, with the entropy being the heat divided by
the equilibrium temperature. \ 

The logic proceeds in the following manner. \ For any given system the
partition function is 
\begin{equation}
Z=Tr\left[ e^{-\beta H}\right]  \label{helm2}
\end{equation}%
where $H$ is the Hamiltonian of the system and the trace is over all of its
possible states. \ Since $U=\left\langle H\right\rangle =Tr\left[ He^{-\beta
H}\right] =-\frac{\partial \ln Z}{\partial \beta }$, the relation 
\begin{equation}
\frac{\partial }{\partial \beta }\left( \beta ^{-1}\ln Z\right) =-\beta
^{-1}\left( \beta ^{-1}\ln Z+U\right)  \label{helm3}
\end{equation}%
is easily seen to hold, from which the identification 
\begin{equation}
\ln Z=-\beta F=S-\beta U  \label{helm4}
\end{equation}%
or 
\begin{equation}
S=\beta \left( U-F\right)  \label{helm5}
\end{equation}%
straightforwardly follows. \ 

\bigskip

To formulate this relation in a gravitational context, one proceeds in an
analogous manner. Consider first the action that gives rise to the Einstein
equations of motion in $d+1$ dimensions, which is 
\begin{equation}
I=I_{B}+I_{\partial B}  \label{act1}
\end{equation}%
where 
\begin{eqnarray}
I_{B} &=&\frac{1}{16\pi G}\int_{{\cal M}}d^{d+1}x\sqrt{-g}\left( R-2\Lambda +%
{\cal L}_{M}\left( \Psi \right) \right)  \label{actB} \\
I_{\partial B} &=&-\frac{1}{8\pi G}\int_{{\cal \partial M}}d^{d}x\sqrt{%
\gamma }\Theta  \label{actdB}
\end{eqnarray}%
${\cal L}_{M}$ is the Lagrangian for matter fields $\Psi $, and a
cosmological constant has been included. The first term in (\ref{actB}) is
the bulk action over the $d+1$ dimensional Manifold ${\cal M}$ with metric $%
g $ . The second term (\ref{actdB}) is a surface term necessary to ensure
that the Euler-Lagrange variation is well-defined, i.e. that one can fix
variations of the metric on the boundary ${\cal \partial M}$ without
constraining variations of metric derivatives. \ This boundary (with induced
metric $\gamma $ and extrinsic curvature $\Theta _{\mu \nu }$) in general
consists of both spacelike and timelike hypersurfaces, and can either be a
boundary of the entire manifold or some submanifold. \ For example, if $%
{\cal \partial M}$ is a boundary of the entire manifold, it will be the
Einstein cylinder at infinity in an asymptotically anti de Sitter (AdS)
spacetime, whereas for an asymptotically de Sitter (dS) spacetime it will be
the union of spatial Euclidean boundaries at early and late times.

One can then construct the path integral 
\begin{equation}
Z=\int D\left[ g\right] D\left[ \Psi \right] e^{-I\left[ g,\Psi \right]
/\hslash }  \label{Zpath}
\end{equation}
by integrating over all metrics and matter fields between some given initial
and final hypersurfaces. \ The integration here is the analog of taking the
trace in eq. (\ref{helm2}). However it is considerably more problematic:
there are continuously infinitely many functional degrees of freedom to
integrate over, there are redundant gauge and diffeomorphism degrees of
freedom that must be eliminated, and the signature of the metric implies
that the integration will not converge.

Remarkably enough one can overcome these difficulties, at least in a
restricted context. \ If the spacetime background were flat, the functional
integration would only be over the matter degrees of freedom. By
analytically continuing the time coordinate $t\rightarrow i\tau $, the path
integral then formally converges. The formalism of finite temperature
quantum field theory \cite{finiteFT}\ then indicates it will become fully
analogous to the expression (\ref{helm2}) provided that the integration over 
$\tau $ in the action is periodic with some period $\beta $. \ Generalizing
to the case of curved spacetimes, a similar set of manipulations can be
performed, provided the class of metrics integrated over is stationary. \
Although the general form of the functional integration is still rather
delicate to perform, to leading order in $\hslash $ one easily obtains 
\[
Z\simeq {\cal N}e^{-I_{\text{cl}}/\hslash } 
\]%
where $I_{\text{cl}}$ is the classical action evaluated on the equations of
motion of the gravity/matter system. Unlike flat-space finite-temperature
field theory, however, the period $\beta $ is no longer arbitrary since
curved Euclideanized manifolds with arbitrary periodic time will typically
have singularities (conical or otherwise) , which in turn imply the
existence of additional matter sources. \ By demanding regularity of the
Euclideanized manifold, such singularities are avoided, and this in turn is
accomplished by restricting the period $\beta $. \ 

The physical interpretation of the preceding formalism is that the class of
regular stationary metrics forms an ensemble of thermodynamic systems at
equilibrium temperature $\beta $. \ Application of eqs. (\ref{helm4},\ref%
{helm5}) then yields 
\begin{equation}
S=\beta {\frak U}-I_{\text{cl}}  \label{entropy}
\end{equation}%
setting $\hslash =1$ and ignoring the normalization factor ${\cal N}$\ ,
which is irrelevant in this semi-classical approximation. The gravitational
entropy $S$ is simply the product of the inverse temperature $\beta $ with
the difference between the total energy ${\frak U}$ and the free energy $%
{\frak F}=I_{\text{cl}}/\beta $. \ 

\bigskip

\section{Conserved\ Charges}

\bigskip

The proposal, then, is that gravitational entropy must be present when there
is a difference between ${\frak U}$ and ${\frak F}$ as defined in equation (%
\ref{entropy}) above. Of course if there is no mismatch between ${\frak U}$
and ${\frak F}$ there will be no gravitational entropy. In this case the
Euclidean manifold will be regular and the period $\beta $ arbitrary, as
with the standard situation in finite-temperature field theory.

Conversely, if there is a difference between ${\frak U}$ and ${\frak F}$
then there must be gravitational heat, and therefore gravitational entropy.
Although the original motivation for obtaining eq. (\ref{entropy}) was to
understand black holes as thermodynamic systems, an extension to
cosmological horizons naturally follows from the formalism \cite{GibH}.
Indeed, as long as the quantities quantities ${\frak U}$, ${\frak F}$ and $%
\beta $ can be computed independently there is no {\it a-priori} reason why
eq. (\ref{entropy}) should not hold in any context.

One situation in which mismatches between ${\frak U}$ and ${\frak F}$ can
occur is when there is a degeneracy in foliating the manifold with foliation
parameter $\tau $. \ Such degeneracies can take place if there is a $U(1)$
isometry (generated by a Killing vector $\xi $) with a fixed-point set. The
existence of any fixed point set makes it impossible to everywhere foliate
the spacetime with surfaces of constant $\tau $, leading to a difference
between the total energy ${\frak U}$ and the free energy\ ${\frak F}$ of the
gravitational system \cite{HHunter,Mismann}. \ 

Evaluation of ${\frak F}$ is tantamount to evaluation of $I_{\text{cl}}$,
and is straightforward for any solution of the Einstein equations. \
Evaluation of $\beta $ can be carried out as outlined above, by demanding
that the Euclidean manifold remain regular at all fixed point sets of the
foliation degeneracy. \ These fixed point sets will have a co-dimension $%
d_{f}\leq d-1$. \ If equality does not hold ( $d_{f}<d-1$) the fixed-point
set is called a ``nut'' \cite{gibh}, in contrast to the ``bolt'' fixed point
sets, whose co-dimension $d_{f}^{\text{bolt}}=d-1$. The regularity
requirement determines $\beta $ in terms of the other parameters in the
metric; if there is more than one fixed-point set then this imposes
additional constraints on these parameters. \ The physical interpretation is
that thermal equilibrium can only be maintained for the set of parameters
obeying these constraints (eg. a black hole in thermal equilibrium with the
cosmological de Sitter temperature). \ \ 

\bigskip

Evaluation of the total energy ${\frak U}$\ is a more delicate matter.
Consider the variation of the action with respect to the metric degrees of
freedom $g_{\mu \nu }$. This will produce two types of terms: an integral
over ${\cal M}$ proportional to the equations of motion, and an integral
over ${\cal \partial M}$ whose integrand $T_{ab}^{\text{eff}}$ is given by
the variation of the action at the boundary with respect to $\gamma ^{ab}$.
It is often referred to as a ``boundary stress-energy'', though it is not
the stress-energy that appears on the right-hand side of the Einstein
equations.\ If the boundary geometry has an isometry generated by a Killing
vector $\xi $, then it is straightforward to show that $T_{ab}^{\text{eff}}$ 
$\xi ^{b}$ is divergenceless with respect to the covariant derivative of the
boundary metric. Writing the boundary metric in the form 
\begin{equation}
\gamma _{ab}d\hat{x}^{a}d\hat{x}^{b}=d\hat{s}^{2}=N_{t}^{2}d\tau ^{2}+\sigma
_{ab}\left( d\varphi ^{a}+N^{a}d\tau \right) \left( d\varphi ^{b}+N^{b}d\tau
\right)  \label{boundmet}
\end{equation}%
where the $\varphi ^{a}$ are coordinates describing closed surfaces $\Sigma $
(distinguished by the foliation parameter $\tau $), and $N_{t}${\Large \ }%
and $N^{a}$\ are the lapse function and shift vector respectively.\ It is
straightforward to show that the quantity 
\begin{equation}
{\frak Q}_{\xi }=\oint_{\Sigma }d^{d-1}S^{a}T_{ab}^{\text{eff}}\xi
^{b}=\oint_{\Sigma }d^{d-1}\varphi \sqrt{\sigma }u^{a}T_{ab}^{\text{eff}}\xi
^{b}  \label{Qxi}
\end{equation}%
is conserved between surfaces of constant $\tau $, whose unit normal is
given by $u^{a}$.\ Physically this means that a collection of observers on
the hypersurface whose metric is $\gamma _{ab}$ all observe the same value
of ${\frak Q}_{\xi }$ provided this surface has an isometry generated by $%
\xi ^{b}$. \ 

If $\xi =\partial /\partial \tau $ then ${\frak Q}_{\xi =\partial /\partial
\tau }$ is identified with the conserved mass/energy ${\frak U}$; if $\xi
_{a}=\partial /\partial \phi ^{a}$ then ${\frak Q}_{\xi _{a}=\partial
/\partial \phi ^{a}}$ is identified with the conserved angular momentum $%
{\frak J}_{a}$ provided $\phi $ is a periodic coordinate associated with $%
\Sigma $. \ 

\bigskip

Computing the gravitational entropy for a given (stationary) gravity/matter
system is then apparently straightforward: (a) calculate $\beta $ by
enforcing regularity of its Euclidean metric, (b) compute its total energy $%
{\frak U}$ using eq. (\ref{Qxi}), (c) compute its classical action $I_{\text{%
cl}}$ from the solution to the Einstein/matter equations, and (d) insert
these into (\ref{entropy}) to obtain the entropy $S$. However the volume of $%
\Sigma $ becomes infinitely large for the entire spacetime, and so neither
the action (\ref{act1}) nor the conserved charges (\ref{Qxi}) are guaranteed
to be finite when evaluated on a solution of the equations of motion. An
obvious response to this situation is to calculate all quantities with
respect to some reference spacetime, interpreted as the ground state of the
system. This can be done by subtracting a term $I_{\text{ref}}\left[ g_{%
\text{ref}},\Psi _{\text{ref}}\right] $ from (\ref{act1}) and embedding $%
\Sigma $ in the background spacetime; conserved quantities then become
differences \ $\Delta {\frak Q}_{\xi }={\frak Q}_{\xi }-{\frak Q}_{\xi }^{%
\text{ref}}$ where the latter term is computed from the reference action. \
Similarly, one must compute $\Delta I=I-I_{\text{ref}}$ , and so the entropy
for a given spacetime in this approach is in general not intrinsically
defined; rather it is given by $\Delta S=S-S_{\text{ref}}$ .

\bigskip

Another procedure that comes to mind is to develop a formalism that does not
require the boundary to contain the entire spacetime. Such a formalism was
originated by Brown and York \cite{BY1}, using a Hamilton-Jacobi type
analysis of the Einstein-Hilbert action. Beginning with a timelike vector
field defining a flow of time, and a timelike foliation of a finite region
of spacetime the action (\ref{act1}) can be decomposed according to this
flow and foliation, yielding natural candidates for conserved charges (such
as energy and momentum) that are defined for a region of finite spatial
extent, i.e. quasilocally.

The advantage of the quasilocal approach is that one can describe
gravitational thermodynamics in regions of finite spatial extent. This is in
more natural accord with our empirical experience of thermodynamics, a
central concept of which is that of a system and a reservoir that are
separated by a partition. Indeed, all physical systems with which we have
had any experience have a finite spatial boundary. \ However the various
thermodynamic variables will depend upon some parameter characterizing the
size of the region -- the radius, for example, if a spherically symmetric
region is chosen as the boundary of the system. \ As this parameter becomes
large, these thermodynamic variables will diverge. \ One can still obtain
results for arbitrarily large regions by embedding the quasilocal boundary
of the region into a reference spacetime as described above, and then
computing all thermodynamic quantities with respect to the referent. \ This
procedure will not affect the first law of thermodynamics for either finite
or infinite regions \cite{BCM,bht}. However the total energy, the entropy,
and all other measurable extensive thermodynamic variables will in general
depend upon the referent background spacetime.

At first this might appear to be a relatively benign modification to the
formalism. \ However it suffers from at least three significant
difficulties. \ First, in an actual physical situation, observers must
contend with the fact that they can only make measurements in the physical
spacetime in which they reside. The reference spacetime is one for which
they have no access -- more simply, there is no guarantee that the ground
state is physically attainable, though it logically exists. Hence there is
no means of empirically checking values relative to some referent. \ One
might hope that this could perhaps be done for all practical purposes, but
even this is not possible. For example one set of observers might surround a
gravitating region and make measurements on a quasilocal surface they
define, whilst a second set of observers agrees to employ the same
quasilocal surface in making their measurements relative to some reference
system. \ The problem is that the second set of observers have no physical
guarantee that the system they are surrounding has the properties required
for it to be a referent since it is not possible to gravitationally screen
this region from other influences, either within or without. \ For example,
dark matter might exist inside it; since by definition such matter can only
be detected gravitationally, there is no way to be certain it is not present
unless measurements are made with respect to a third referent, plunging the
observers into a situation of infinite regress.

These issues could be avoided, at least conceptually, if the reference
spacetime could be regarded purely as a mathematical artifact that all
observers agreed to refer their measurements to. Unfortunately -- and this
is the second difficulty -- the choice of reference spacetime is not always
unique \cite{CCM}. \ In principle there can be several different choices one
could make for such a referent, all of which are physically reasonable, but
each of which yields different answers for the quantities in question. \ The
third difficulty is that observers can choose quasilocal system boundaries
for which the referent does not even exist, since it is not always possible
to embed a boundary with a given induced metric into the reference
background. \ This problem is not confined to esoteric examples; it occurs
even for the simple case of a Kerr black hole, where it has been shown that
the embedding problem forms a serious obstruction towards calculating the
subtraction energy \cite{Martinez}. \ 

\section{Boundary Counterterms}

It would be desirable, then, to have a formalism that avoids the use of a
reference spacetime entirely, defining all thermodynamic variables with
respect to quantities intrinsic to the spacetime. \ This would resolve the
measurement problem since no referent spacetime exists, as well as the
embedding problem, since there is no need to embed. \ One way of doing this
is to introduce additional terms in the action that depend only on curvature
invariants that are functionals of the intrinsic boundary geometry. \ Such
terms cannot alter the equations of motion and, since they are divergent,
offer the possibility of removing divergences that arise in the action (\ref%
{act1}) for arbitrarily large spacetime regions. Although there exist
infinitely many curvature invariants for a given boundary metric, for
dimensional reasons only finitely many of them will be non-zero in the limit
the boundary contains the entire spacetime. \ Consequently one need choose
only a finite number of coefficients to cancel the divergences of a given
spacetime.

In this approach, then, all observers agree to use an action of the form 
\begin{equation}
I=I_{B}+I_{\partial B}+I_{\text{ct}}\left( \gamma \right)  \label{fullact}
\end{equation}
where the first two terms are given by (\ref{actB},\ref{actdB}) above and $%
I_{\text{ct}}\left( \gamma \right) $ depends on the intrinsic geometry of
the boundary. This latter term acts as kind of ``counterterm'' that cancels
out the divergences in the action and conserved charges from the first two
terms. \ Indeed, this approach \cite{henhybala} was inspired by the
conjectured AdS/CFT correspondence \cite{Maldecena}, which states that the
partition function of any field theory on $AdS_{d+1}$ is identified with the
generating functional $Z_{CFT}$ of a conformal field theory on its boundary $%
\partial {\cal M}{_{d}}$ at infinity \cite{FWG} 
\begin{equation}
\left\langle \exp \left( \int_{\partial {\cal M}{_{d}}}d^{d}x\sqrt{g}{\cal O}%
\phi _{0}\right) \right\rangle =Z_{CFT}[\phi _{0}]\equiv Z_{AdS}[\phi
_{0}]=\int_{{\phi _{o}}}{\cal D}\phi \text{ }e^{-S(\phi )}\text{ }
\label{PAR}
\end{equation}
where $\phi _{0}$ is the finite field defined on the boundary of $AdS_{d+1}$%
, the integration is over the field configurations $\phi $ that approach $%
\phi _{0}$ when one goes from the bulk of $AdS_{d+1}$ to its boundary, and $%
{\cal O}$ is a quasi-primary conformal operator on $\partial {\cal M}{_{d}}$
. While there is no proof at present of this conjecture, there is
considerable circumstantial evidence to support it \cite{Adscftex} . For
example it has been explicated for a free massive scalar field and a free $%
U(1)$ gauge theory, as well as having been partially confirmed for an
interacting massive scalar, a free massive spinor and interacting
scalar-spinor fields, as well as for classical gravity and type-IIB string
theory. In all these cases, the exact partition function (\ref{PAR}) is
given by the exponential of the action evaluated for a classical field
configuration which solves the classical equations of motion, and explicit
calculations show that the evaluated partition function is equal to the
generating functional of some conformal field theory with a quasi-primary
operator of a certain conformal weight.

Quantum field theories in general contain counterterms, and so it is natural
from the AdS/CFT viewpoint to append the boundary term $I_{ct}$\ to the
action as in eq. (\ref{fullact}). \ However there remains the issue of
uniqueness, since it is logically admissible that the coefficients must be
chosen differently for each spacetime under consideration, or alternatively,
that there exist two different choices of coefficients for a given spacetime
that give rise to different finite results. \ Remarkably enough neither of
these situations arises for asymptotically AdS spacetimes; rather the
boundary counterterm action is universal, being composed of a unique linear
combination of curvature invariants that cancel the divergences that arise
in the limit the boundary contains the full spacetime. \ This was first
observed for the full range of type-D asymptotically AdS spacetimes,
including Schwarzchild-AdS, Kerr-AdS, Taub-NUT-AdS, Taub-bolt-AdS, and\
Taub-bolt-Kerr-AdS \cite{Mismann,EJM,nutkerr}; shortly afterward a
straightforward algorithm was constructed for generating it \cite{KLS}. \
The procedure involves solving the Einstein equations (written in
Gauss-Codacci form) in terms of the extrinsic curvature functional $\Pi
_{ab}=\Theta _{ab}-\Theta \gamma _{ab}$ of the boundary ${\cal \partial M}$
and its normal derivatives to obtain the divergent parts. The algorithm
works because all divergent parts can be expressed in terms of intrinsic
boundary data and so do not depend on normal derivatives \cite{Feffgraham}.
\ Explicitly, one writes the divergent part $\tilde{\Pi}_{ab}$ as a power
series in the inverse cosmological constant, then\ covariantly isolates the
entire divergent structure for any given boundary dimension $d$. Varying the
boundary metric under a Weyl transformation, it is straightforward to show
that the trace $\tilde{\Pi}$ is proportional to the divergent boundary
counterterm Lagrangian.

The actual form of $I_{\text{ct}}\left( \gamma \right) $ has only been
explicitly computed up to $9$ dimensions \cite{saurya} 
\begin{eqnarray}
I_{\text{ct}} &=&\int \sqrt{-\gamma }\left[ -\frac{d-1}{\ell }-\frac{\ell 
{\sf \Theta }\left( d-3\right) }{2(d-2)}{\sf R}-\frac{\ell ^{3}{\sf \Theta }%
\left( d-5\right) }{2(d-2)^{2}(d-4)}\left( {\sf R}^{ab}{\sf R}_{ab}-\frac{d-1%
}{4(d-2)}{\sf R}^{2}\right) \right.  \nonumber \\
&&+\frac{\ell ^{5}{\sf \Theta }\left( d-7\right) }{(d-2)^{3}(d-4)(d-6)}%
\left( \frac{3d+2}{4(d-1)}{\sf RR}^{ab}{\sf R}_{ab}-\frac{d(d+2)}{16(d-1)^{2}%
}{\sf R}^{3}\right.  \nonumber \\
&&\text{ \ \ }\left. \text{\ \ \ \ \ }\left. -2{\sf R}^{ab}{\sf R}^{cd}{\sf R%
}_{acbd}-\frac{d}{4(d-1)}\nabla _{a}{\sf R}\nabla ^{a}{\sf R}+\nabla ^{c}%
{\sf R}^{ab}\nabla _{c}{\sf R}_{ab}\right) \right]  \label{Lct2}
\end{eqnarray}%
where ${\sf R}$ denotes the curvature tensor (and its contractions) of the
boundary, and the step function ${\sf \Theta }(x)$ is equal to zero unless $%
x\geq 0$ in which case it equals unity. The quantity $\ell
^{2}=-d(d-1)/2\Lambda $, and diverges in the flat-space limit. \ 

\bigskip

It was recently shown that this approach can be generalized to
asymptotically de Sitter spacetimes \cite{bala,baladesit}. \ The relevant
boundaries ${\cal \partial M}^{\pm }$ are now spatial Euclidean boundaries
at early and late times. The boundary and counterterm actions become 
\begin{equation}
I_{\partial B}=\frac{\beta }{16\pi G}\int_{{\cal \partial M}^{-}}^{{\cal %
\partial M}^{+}}d^{d}x\sqrt{\gamma ^{\pm }}\Theta ^{\pm }  \label{actbound}
\end{equation}%
\begin{equation}
I_{ct}=\frac{1}{16\pi G}\int_{{\cal \partial M}^{-}}^{{\cal \partial M}%
^{+}}d^{d}x\sqrt{\gamma ^{\pm }}{\cal L}_{ct}^{\pm }  \label{counter}
\end{equation}%
It has been shown that the divergences of asymptotically de Sitter
spacetimes are independent of the boundary normal, and so depend only on
intrinsic boundary data \cite{Mozzola}. Using this information a counterterm
algorithm similar to that for the AdS case can be constructed. The result is 
\begin{eqnarray}
{\cal L}_{ct} &=&\left( -\frac{d-1}{\ell }+\frac{\ell {\sf \Theta }\left(
d-3\right) }{2(d-2)}{\sf R}\right) -\frac{\ell ^{3}{\sf \Theta }\left(
d-5\right) }{2(d-2)^{2}(d-4)}\left( {\sf R}^{ab}{\sf R}_{ab}-\frac{d}{4(d-1)}%
{\sf R}^{2}\right)  \label{countergen} \\
&&-\frac{\ell ^{5}{\sf \Theta }\left( d-7\right) }{(d-2)^{3}(d-4)(d-6)}%
\left( \frac{3d+2}{4(d-1)}{\sf RR}^{ab}{\sf R}_{ab}-\frac{d(d+2)}{16(d-1)^{2}%
}{\sf R}^{3}-2{\sf R}^{ab}{\sf R}^{cd}{\sf R}_{acbd}\right.  \nonumber \\
&&\text{ \ \ \ \ \ \ \ \ \ \ \ \ \ \ \ \ \ \ \ \ \ \ \ \ \ \ \ \ \ \ \ }%
\left. -\frac{d}{4(d-1)}\nabla _{a}{\sf R}\nabla ^{a}{\sf R}+\nabla ^{c}{\sf %
R}^{ab}\nabla _{c}{\sf R}_{ab}\right)  \nonumber
\end{eqnarray}

Turning next to a consideration of conserved charges, if the boundary
geometry has an isometry generated by a Killing vector $\xi ^{\mu }$, then a
conserved ${\frak Q}_{\xi }$ can be defined as in (\ref{Qxi}). However in
the asymptotically de Sitter context its physical interpretation is somewhat
different: a collection of observers on the hypersurface whose metric is $%
\gamma _{ab}$ would all observe the same value of ${\frak Q}$ provided this
surface had an isometry generated by $\xi ^{b}$. \ Although the charge
changes with the cosmological time $\tau $, a collection of observers that
defined a surface $\Sigma $ would find that the value of ${\frak Q}$ they
would measure is the same as that of other observers collectively relocated
elsewhere on the spacelike \ surface ${\cal \partial M}$. Although this
surface does not enclose anything, it does not matter -- the conserved
charge is associated only with $\Sigma $, independently of either the
structure or existence of the interior of $\Sigma $ \cite{ibooth}.

{\bf \bigskip }

\section{Examples}

Consider an evaluation of the energy of Schwarzschild-AdS spacetime with
static slicing, that is, 
\begin{equation}
ds^{2}=-f(r)dt^{2}+\frac{1}{f(r)}dr^{2}+r^{2}\sigma _{ij}dx^{i}dx^{j},
\label{SADS}
\end{equation}
where $\sigma _{ij}$ is a metric of a unit $\left( d-1\right) $-dimensional
sphere, plane or (compact)\ hyperboloid for $k=1,0,-1$, respectively, and $%
f(r)=k-2m/r^{d-2}+r^{2}/\ell ^{2}$. \ \ The Euclideanized metric is obtained
by setting $t\rightarrow i\tau $ 
\begin{equation}
ds^{2}=f(r)d\tau ^{2}+\frac{1}{f(r)}dr^{2}+r^{2}\sigma _{ij}dx^{i}dx^{j},
\label{SADSE}
\end{equation}
upon which it is clear that wherever $f(r)$ vanishes there is a degeneracy
in foliating the metric with surfaces of constant \ $\tau $. The metric will
thus have a conical singularity in the $\left( \tau ,r\right) $\ section
unless $\tau $ is appropriately periodically identified. \ Writing $r=\frac{1%
}{4}f_{+}^{\prime }\varepsilon ^{2}+r_{+}$, where $f(r_{+})=0$ and \ $%
f_{+}^{\prime }=$ $\left. \frac{df}{dr}\right| _{r=r_{+}}$, the metric
becomes 
\begin{equation}
ds^{2}=\frac{1}{4}\left( f_{+}^{\prime }\varepsilon \right) ^{2}d\tau
^{2}+d\varepsilon ^{2}+r^{2}\sigma _{ij}dx^{i}dx^{j},  \label{conical}
\end{equation}
and so the period must be 
\begin{equation}
\beta =\left| \frac{4\pi }{f_{+}^{\prime }}\right| =\frac{2\pi \ell
^{2}r_{+}^{d-1}}{r_{+}^{d}+(d-2)m\ell ^{2}}  \label{betasads}
\end{equation}
in order to ensure regularity in the region $r\geq r_{+}$. \ 

Consider a set of observers that place themselves in a collective position
of radial symmetry; this is tantamount to enclosing the system with a
surface at constant $R$. \ \ This surface has a spacelike unit normal $%
n^{\alpha }=\hat{r}^{\alpha }=\left( \partial /\partial r\right) ^{\alpha }$
and extrinsic curvature 
\begin{equation}
\Theta _{tt}=-\frac{f^{\prime }\left( R\right) }{2}\sqrt{f(R)}~~~~~~~\Theta
_{ij}=R\sqrt{f(R)}\sigma _{ij}  \label{EXSADS}
\end{equation}%
and so 
\begin{equation}
\left( \Theta _{\mu \nu }-\Theta \gamma _{\mu \nu }\right) u^{\mu }\xi ^{\nu
}=-\left( d-1\right) \frac{f(R)}{R}  \label{Edivsads}
\end{equation}%
\bigskip is the energy density in the absence of the counterterm
contributions. \ When integrated over the surface as in (\ref{Qxi}), this
quantity diverges as $R\rightarrow \infty $ . \ \ However inclusion of the
counterterm contributions\ from the variation of (\ref{Lct2}) yields {\bf \ }
\begin{eqnarray}
T_{tt}^{\text{eff}} &=&\frac{\left( d-1\right) f(R)}{8\pi G_{n}\ell }\biggl\{%
-\frac{\ell \sqrt{f(R)}}{R}+\left[ {\sf \Theta }\left( d-1\right) +\frac{%
k\ell ^{2}{\sf \Theta }\left( d-3\right) }{2R^{2}}\right]  \nonumber \\
&&-\frac{k^{2}\ell ^{4}{\sf \Theta }\left( d-5\right) }{8R^{4}}+\frac{%
k^{3}\ell ^{6}{\sf \Theta }\left( d-7\right) }{16R^{6}}+\cdots \biggr\} 
\nonumber \\
&=&\frac{\left( d-1\right) f(R)}{8\pi G_{n}\ell }\left[ \left. \sqrt{1+\frac{%
k\ell ^{2}}{R^{2}}}\right\rfloor _{d}-\sqrt{1-\frac{2m\ell ^{2}}{R^{d}}+%
\frac{k\ell ^{2}}{R^{2}}}\right]  \label{Teffsads}
\end{eqnarray}%
explicitly showing all terms up to nine dimensions ($d=8$). \ \ 

The notation for the first square root means that it is to be understood as
a power series that is truncated so that only $d$ terms are retained. \ This
has some interesting implications for large $R$. \ If there were no
truncation of the series, then the leading term in (\ref{Teffsads}) $\sim
mR^{1-d}$, independent of the value of $k$. \ For odd $d$, this remains true
even when the series is truncated. \ However for even $d$ there is
contribution proportional to $k^{d/2}R^{1-d}$ that is not cancelled because
of the truncation.

Consequently the total energy is 
\begin{eqnarray}
{\frak U}_{{\rm AdS}}^{\left( d,k\right) } &=&\int_{\sigma }d^{d-1}x\sqrt{%
\sigma }T_{\mu \nu }^{\text{eff}}u^{\mu }\xi ^{\nu }=\frac{\left( d-1\right)
R^{d-1}V_{d-1}\sqrt{f(R)}}{8\pi G_{n}\ell }\left[ \left. \sqrt{1+\frac{k\ell
^{2}}{R^{2}}}\right\rfloor _{d}-\frac{\ell }{R}\sqrt{f(R)}\right]  \nonumber
\\
&\longrightarrow &V_{d-1}\frac{\left( d-1\right) }{8\pi G_{d}}\left[ m+\frac{%
\Gamma \left( \frac{2p-3}{2}\right) \ell ^{2p-4}\left( -k\right) ^{(p-1)}}{2%
\sqrt{\pi }\Gamma \left( p\right) }\delta _{2p,d}\right]  \label{Utotalsads}
\end{eqnarray}
in the limit $R\rightarrow \infty $, where $\xi ^{\nu }$ is the timelike
Killing vector $\partial /\partial t$ and $V_{d-1}$ is the volume of the
surface of the compact $\left( d-1\right) $-dimensional space (for $k=1$,
see \cite{saurya}). If $p=\frac{d}{2}$ is a positive integer, the second
term remains. This extra term is interpreted as a Casimir energy in the
context of the AdS/CFT correspondence conjecture. Its sign depends on the
dimensionality and the value of $k$. Note that slices with $k=0$ have ${\Bbb %
R}^{4}$ topology and are flat, and the Casimir energy vanishes. This is
compatible with the features of quantum field theory in curved spacetime.

The action is straightforwardly evaluated. From (\ref{actB}) the bulk
component is 
\begin{eqnarray}
I_{B} &=&-\frac{1}{16\pi G_{d}}\int_{{\cal M}}d^{d+1}x\sqrt{-g}\left( -\frac{%
d\left( d+1\right) }{\ell ^{2}}+\frac{d\left( d-1\right) }{\ell ^{2}}\right)
\nonumber \\
&=&\frac{d}{8\pi G_{d}\ell ^{2}}\int_{{\cal M}_{d}}d^{d}\hat{x}\sqrt{\sigma }%
\int_{r_{+}}^{R}drr^{d-1}  \nonumber \\
&=&\frac{\beta V_{d-1}}{8\pi G_{d}\ell ^{2}}\left( R^{d}-r_{+}^{d}\right)
\label{Ibulksads}
\end{eqnarray}%
whereas the boundary action (\ref{actdB}) is given by 
\begin{equation}
I_{\partial B}=-\frac{1}{8\pi G_{d}}\int_{{\cal \partial M}}d^{d}x\sqrt{%
\gamma }\Theta =-\frac{\beta V_{d-1}R^{d-1}}{8\pi G_{d}}\left[ \frac{%
f^{\prime }\left( R\right) }{2}+\left( d-1\right) \frac{f\left( R\right) }{R}%
\right]  \label{Ibndsads}
\end{equation}%
and the counterterm action (\ref{Lct2}) is 
\begin{eqnarray}
I_{\text{ct}} &=&\frac{1}{8\pi G_{d}}\int_{{\cal \partial M}}d^{d}x\sqrt{%
\gamma }\left[ \left( \frac{d-1}{\ell }+\frac{k\ell {\sf \Theta }\left(
d-3\right) }{2R^{2}}\left( d-1\right) \right) +\frac{k^{2}\ell ^{3}{\sf %
\Theta }\left( d-5\right) }{8R^{4}}\left( d-1\right) +\frac{k^{3}\ell ^{5}%
{\sf \Theta }\left( d-7\right) }{16R^{6}}\left( d-1\right) \right]  \nonumber
\\
&=&\frac{\left( d-1\right) \beta V_{d-1}R^{d-1}}{8\pi G_{d}\ell }\sqrt{%
f\left( R\right) }\left. \sqrt{1+\frac{k\ell ^{2}}{R^{2}}}\right\rfloor _{d}
\label{Ictsads}
\end{eqnarray}%
yielding 
\begin{equation}
I=\frac{\left( d-1\right) \beta V_{d-1}R^{d-1}}{8\pi G_{d}\ell }\left( \sqrt{%
f\left( R\right) }\left. \sqrt{1+\frac{k\ell ^{2}}{R^{2}}}\right\rfloor _{d}-%
\frac{\ell f\left( R\right) }{R}\right) -\frac{\beta V_{d-1}\left(
r_{+}^{d}+m\ell ^{2}\left( d-2\right) \right) }{8\pi G_{d}\ell ^{2}}
\label{Itotsads}
\end{equation}%
for the total Euclidean action. \ This becomes 
\begin{equation}
I=\frac{\beta V_{d-1}}{8\pi G_{d}}\left[ m-\frac{r_{+}^{d}}{\ell ^{2}}%
+\left( d-1\right) \frac{\Gamma \left( \frac{2p-3}{2}\right) \ell
^{2p-4}\left( -k\right) ^{(p-1)}}{2\sqrt{\pi }\Gamma \left( p\right) }\delta
_{2p,d}\right]  \label{Itotsimpsads}
\end{equation}%
in the $R\rightarrow \infty $ limit

Finally, using eq. (\ref{entropy}) the gravitational entropy is given by 
\begin{equation}
S_{d}=\frac{\left( r_{+}^{d}+(d-2)m\ell ^{2}\right) \beta V_{d-1}}{8\pi
G_{d}\ell ^{2}}=\frac{r_{+}^{d-1}V_{d-1}}{4G_{d}}=\frac{1}{4G}A_{d-1}
\label{entsads}
\end{equation}%
which is the familiar entropy/area relation for the $\left( d+1\right) $%
-dimensional class of Schwarzchild anti de Sitter metrics, with \ $A_{d-1}$
the area of the event horizon. \ 

The preceding derivation is deceptively simple. Since the counterterm action
(\ref{Lct2}) is universal and unique (at least for asymptotically AdS
spacetimes) it would suggest that the results (\ref{Utotalsads},\ref%
{Itotsimpsads}) are also unique, but such is not the case. \ Under a
coordinate transformation, the metric (\ref{SADS}) becomes for $k=-1$ 
\begin{equation}
ds^{2}=g^{2}d{\frak r}^{2}+a^{2}\omega ^{4/(d-2)}(-d{\frak t}^{2}+{\frak t}%
^{2}\sigma _{ij}dx^{i}dx^{j}),  \label{sitadmet}
\end{equation}%
with $a(r)=e^{-{\frak r}/\ell }$, $g=2/\omega -1$ and $\omega =1-m/2[a(%
{\frak r}){\frak t}]^{d-2}$, and an analogous calculation for a collection
of observers situated at constant ${\frak r}={\frak R}$ yields \ 
\begin{equation}
{\frak U}_{{\rm AdS/CFS}}^{\left( d,k=-1\right) }=V_{d-1}\frac{m\left(
d-1\right) }{4\pi G_{d}}  \label{ActAdS}
\end{equation}%
for the total energy and action as ${\frak R}\rightarrow \infty $. \ The
total energy is now given by a completely different expression for $d=$even,
indicating a clear dependence of these quantities on the foliation of the
spacetime. \ 

\bigskip

As a second example, consider the following metric 
\begin{equation}
ds^{2}=-V(r)\left( dt+2n\cos (\theta )d\phi \right) +\frac{dr^{2}}{V(r)}%
+(r^{2}+n^{2})\left( d\theta ^{2}+\sin ^{2}(\theta )d\phi ^{2}\right)
\label{4dmetric}
\end{equation}%
which describes an asymptotically anti de Sitter spacetime with a non-zero
NUT\ charge $N$ in four dimensions. \ Here 
\begin{equation}
V(r)=\frac{r^{2}-n^{2}-2mr+\ell ^{-2}(r^{4}+6n^{2}r^{2}-3n^{4})}{r^{2}+n^{2}}
\label{Fr4d}
\end{equation}%
This spacetime has singularities along the $\theta =\left( 0,\pi \right) $
axes, as is easily seen by computing $\left( \nabla t\right) ^{2}$ at these
locations. Such ``Misner string'' singularities are the analog of Dirac
string singularities that accompany magnetic monopoles, and can be removed
by appropriate periodic identification of the time coordinate $t$ \cite%
{Misner}. \ 

\bigskip

\ Here again we see an example of a breakdown in foliating the spacetime by
a family of surfaces of constant time. In the Euclidean regime this
situation can occur in $d$-dimensions if the topology of the Euclidean
spacetime is not trivial -- specifically when the (Euclidean) timelike
Killing vector $\xi =\partial /\partial \tau $\ that generates the U(1)
isometry group has a fixed point set of even co-dimension. If this
co-dimension is $\left( d-1\right) $ then the usual relationship between
area and entropy holds. However if the co-dimension is smaller than this the
relationship between area and entropy is generalized as we shall see.

Setting $t\rightarrow i\tau $ and $n\rightarrow iN$ yields 
\begin{equation}
ds^{2}=F(r)\left( d\tau +2N\cos (\theta )d\phi \right) +\frac{dr^{2}}{F(r)}%
+(r^{2}-N^{2})\left( d\theta ^{2}+\sin ^{2}(\theta )d\phi ^{2}\right)
\label{TnutEuc}
\end{equation}
with 
\begin{equation}
F(r)=\frac{r^{2}+N^{2}-2mr+\ell ^{-2}(r^{4}-6N^{2}r^{2}-3N^{4})}{r^{2}-N^{2}}
\label{Feuc}
\end{equation}
which describes the Euclidean section of this spacetime. Computing the total
Euclidean action from (\ref{fullact}) gives

\begin{equation}
I_{4}=\frac{\beta }{2\ell ^{2}}\left( \ell ^{2}m+3N^{2}r_{+}-r_{+}^{3}\right)
\label{4ditot}
\end{equation}
where $r_{+}$ is a root of $F(r)$ and $\beta $ is the period of $\tau $,
given in four dimensions by 
\begin{equation}
\beta =8\pi N  \label{bet4}
\end{equation}
and determined by demanding regularity of the manifold so that the
singularities at $\theta =0,\pi $ are coordinate artifacts.

There is an additional regularity criterion to be satisfied, namely the
absence of conical singularities at the roots of the function $F(r)$. The
argument is similar to that which gave rise to eq. (\ref{betasads}) \ and
yields 
\begin{equation}
\beta =\left| \frac{4\pi }{F^{\prime }\left( r_{+}\right) }\right| =8\pi N
\label{bet4a}
\end{equation}%
where $F(r=r_{+})=0$; the second equality follows by demanding consistency
with eq. (\ref{bet4}). \ There are only two solutions to eq. (\ref{bet4a}),
one where $r_{+}=N$ and one where $r_{+}=r_{b}>N$, referred to respectively
as the NUT and Bolt solutions. The latter solutions have $2$-dimensional
fixed point sets, whereas the former have $0$-dimensional fixed point sets.
\ When the NUT charge is nonzero, the entropy of a given spacetime includes
not only the entropies of the $2$-dimensional bolts, but also those of the
Misner strings, and (as noted by Hawking and Hunter \cite{HHunter}) should
contribute to the gravitational entropy.

\bigskip

Solving (\ref{bet4a}) with $r_{+}=r_{b}>N$ gives two possible solutions 
\begin{equation}
r_{b\pm }=\frac{\ell ^{2}\pm \sqrt{\ell ^{4}-48N^{2}\ell ^{2}+144N^{4}}}{12N}
\label{rbolt4}
\end{equation}
where 
\begin{equation}
N\leq \frac{(3\sqrt{2}-\sqrt{6})\ell }{12}=N_{max}  \label{Nlimit4}
\end{equation}
so that $r_{b}$ remains real.

Restricting eq. (\ref{4ditot}) to the NUT and Bolt cases yields 
\begin{eqnarray}
I_{NUT} &=&\frac{4\pi N^{2}(\ell ^{2}-2N^{2})}{\ell ^{2}}  \label{INUT4d} \\
I_{Bolt} &=&\frac{-\pi (r_{b}^{4}-\ell ^{2}r_{b}^{2}+N^{2}(3N^{2}-\ell ^{2}))%
}{3r_{b}^{2}-3N^{2}+\ell ^{2}}  \label{IBolt4d}
\end{eqnarray}
with the conserved total energy being 
\begin{eqnarray}
{\frak U}_{\text{NUT}} &=&\frac{N(\ell ^{2}-4N^{2})}{\ell ^{2}}  \label{UNUT}
\\
{\frak U}_{\text{Bolt}} &=&\frac{r_{b}^{4}+(\ell
^{2}-6N^{2})r_{b}^{2}+N^{2}(\ell ^{2}-3N^{2})}{2\ell ^{2}r_{b}}
\label{Ubolt}
\end{eqnarray}
for each case. \ The entropies then follow from the relation (\ref{entropy}) 
\begin{eqnarray}
S_{\text{NUT}} &=&\frac{4\pi N^{2}(\ell ^{2}-6N^{2})}{\ell ^{2}}
\label{SNUT} \\
S_{\text{Bolt}} &=&\pi \frac{3r_{b}^{4}+(\ell
^{2}-12N^{2})r_{b}^{2}+N^{2}(\ell ^{2}-3N^{2})}{3r_{b}^{2}-3N^{2}+\ell ^{2}}
\label{Sbolt}
\end{eqnarray}
and are no longer given by the areas of the respective event horizons.

The counterterm prescription affords an intrinsic thermodynamic description
of the spacetimes described by (\ref{4dmetric}), but with rather strange
results \cite{Mismann,EJM,nutkerr}. From eq. (\ref{SNUT}) we see that the
entropy becomes negative for $N>{\frac{\ell }{\sqrt{6}}}$. A further
calculation shows that the specific heat becomes negative for $N<{\frac{\ell 
}{\sqrt{12}}}$. Hence there are no thermally stable solutions outside of the
range 
\begin{equation}
\frac{\ell }{\sqrt{12}}\leq N\leq \frac{\ell }{\sqrt{6}}  \label{rangeN}
\end{equation}%
A similar analysis of the bolt case shows that the upper branch solutions ($%
r_{b}=r_{b+}$) are thermally stable, whereas the lower branch solutions ($%
r_{b}=r_{b-}$) are thermally unstable.

A forthcoming analysis \cite{Ricklorenzo} of higher-dimensional versions of
the spacetimes described by (\ref{4dmetric}) indicates that their
thermodynamic behaviour is similar for every dimension $d=4k$. \ However for 
$d=4k+2$, it can be shown that nowhere are the entropy and specific heat
positive for the same values of $N$.

Corroborating evidence for the preceding results is given by demonstrating
that a Noether-charge interpretation of this more general notion of
gravitational entropy exists \cite{NUTNoether}. \ 

\bigskip

As a final example, consider $d+1$ dimensional Schwarzschild-deSitter
spacetime, with metric 
\begin{equation}
ds^{2}=-N(r)dt^{2}+\frac{dr^{2}}{N(r)}+r^{2}d\hat{\Omega}_{d-1}^{2}
\label{SdSmet}
\end{equation}%
where 
\begin{equation}
N(r)=1-\frac{2m}{r^{d-2}}-\frac{r^{2}}{\ell ^{2}}
\end{equation}%
and $d\hat{\Omega}_{d-1}^{2}$ denotes the metric on the unit sphere $%
S^{d-1}. $ There is a black hole with event horizon at $r=r_{H}$ and
cosmological horizon at $r=r_{C}>r_{H}$ (located at the roots of $N(r)$)
provided that 
\begin{equation}
0<m\leq m_{N}=\frac{\ell ^{d-2}}{d}(\frac{d-2}{d})^{\frac{d-2}{2}}
\end{equation}%
where saturation of the inequality yields a spacetime referred to as the
Nariai solution. For $m>m_{N},$ the metric (\ref{SdSmet}) describes a naked
singularity in an asymptotically dS spacetime. Hence regularity requirements
yield an upper limit to the mass of the SdS black hole.

Working outside of the cosmological horizon, where $N(r)<0$, \ the metric
can be rewritten as 
\begin{equation}
ds^{2}=-f(\tau )d\tau ^{2}+\frac{dt^{2}}{f(\tau )}+\tau ^{2}d\tilde{\Omega}%
_{d-1}^{2}
\end{equation}
where $r=\tau $ and 
\begin{equation}
f(\tau )=\left( \frac{\tau ^{2}}{\ell ^{2}}+\frac{2m}{\tau ^{d-2}}-1\right)
^{-1}
\end{equation}
where the cosmological horizon is at $\tau _{+}$, defined to be the largest
root of $\ \left[ f(\tau _{+})\right] ^{-1}=0$. Working in this ``upper
patch'' outside of the cosmological horizon, the bulk action is now 
\begin{eqnarray}
I_{B} &=&\frac{d}{8\pi G\ell ^{2}}\int d^{d}x\int_{\tau _{+}}^{\tau }d\tau 
\sqrt{f}\sqrt{h}=\frac{d}{8\pi G\ell ^{2}}\int dtd^{d-1}\hat{x}\sqrt{\sigma }%
\int_{\tau _{+}}^{\tau }d\tau \tau ^{d-1}  \nonumber \\
&=&\frac{V_{d-1}^{t}}{8\pi G\ell ^{2}}\left( \tau ^{d}-\tau _{+}^{d}\right)
\end{eqnarray}
where the integration is from the cosmological horizon out to some fixed $%
\tau $ that will be sent to infinity. Here $V_{d-1}^{t}=\int dtd^{d-1}\hat{x}%
\sqrt{\sigma }$, with $\sigma ^{ab}$ the metric on the unit $(d-1)$-sphere.

\bigskip

Including the remaining boundary contributions to the action yields 
\begin{eqnarray}
I &=&\frac{V_{d-1}^{t}\tau ^{d-1}}{8\pi G}\left[ \frac{1}{\ell ^{2}}(\tau -%
\frac{\tau _{+}^{d}}{\tau ^{d-1}})-\frac{1}{2f}\left( -\frac{f^{\prime }}{f}+%
\frac{2\left( d-1\right) }{\tau }\right) \right.  \nonumber \\
&&-\frac{\left( d-1\right) }{\sqrt{f}}\left. \left\{ \left( -\frac{1}{\ell }+%
\frac{\ell \Theta \left( d-3\right) }{2\tau ^{2}}\right) +\frac{\ell
^{3}\Theta \left( d-5\right) }{8\tau ^{4}}+\frac{\ell ^{5}\Theta \left(
d-7\right) }{16\tau ^{6}}\right\} \right]  \label{actdesit}
\end{eqnarray}
for the total action in this patch. In the large $\tau $-limit this becomes
\ 
\begin{equation}
I=\frac{\beta V_{d-1}}{4\pi G_{d}}\left[ m+\frac{\tau _{+}^{d}}{2\ell ^{2}}%
-\left( d-1\right) \frac{\Gamma \left( \frac{2p-1}{2}\right) \ell ^{2p-2}}{2%
\sqrt{\pi }\Gamma \left( p+1\right) }\delta _{2p,d}\right]  \label{actdesitd}
\end{equation}
The total energy is 
\begin{equation}
{\frak U}_{{\rm dS}}^{\left( n\right) }=V_{d-1}\frac{\left( d-1\right) }{%
4\pi G_{n}}\left[ -m+\frac{\Gamma \left( \frac{2p-1}{2}\right) \ell ^{2p-2}}{%
2\sqrt{\pi }\Gamma \left( p+1\right) }\delta _{2p,d}\right]  \label{MSDS}
\end{equation}

For odd values of $d$, the total energy \ ${\frak U}$ is negative whereas
for even values of $\ d$ it is positive. As the mass parameter $m$
increases, this positive value decreases, approaching its minimum at the
Nariai limit.\ Setting $m=0$ gives the total energy of dS spacetime in
different dimensionalities.

The volumes $V_{d-1}$ are in general divergent, since the $t$-coordinate is
of infinite range. However since $\partial /\partial t$ is a Killing vector,
it is tempting to periodically identify it; analytically continuing $%
t\rightarrow it$, yields a metric of signature $(-2,d-1)$. \ The section
with signature $\left( -,-\right) $ (described by the $(t,\tau )$
coordinates) must have a periodic identification of the $t$-coordinate with
period 
\begin{equation}
\beta _{H}=\left| \frac{4\pi }{(-N^{\prime }(r))}\right| _{r=\tau
_{+}}=\left| \frac{-4\pi f^{\prime }(\tau )}{f^{2}}\right| _{\tau =\tau _{+}}
\label{betaH}
\end{equation}
so that there is no conical singularity at $\tau =\tau _{+}$ .

This quantity $\beta _{H}$ is the analogue of the Hawking temperature
outside of the cosmological horizon. \ Proceeding further, the entropy, as
defined by relation (\ref{entropy}) becomes 
\begin{equation}
S_{d}=\frac{\left( \tau _{+}^{d}-2(d-2)m\ell ^{2}\right) \beta _{H}V_{d-1}}{%
8\pi G\ell ^{2}}  \label{sdsentropy}
\end{equation}
Remarkably enough these entropies are always positive, since $\tau
_{+}^{d}>2(d-2)m\ell ^{2}$ so long as $m<m_{N}$. \ For example, for $d=2$,\ $%
\tau _{+}=\ell \sqrt{1-2m}$and $\beta _{H}=2\pi \ell ^{2}/\tau _{+}$,
yielding 
\begin{equation}
S^{d=2}=\frac{\tau _{+}V_{1}}{4G}=\frac{\pi \ell \sqrt{1-2m}}{2G}
\label{d2entropy}
\end{equation}

The generalized entropy (\ref{sdsentropy}) is a monotonically decreasing
function of the mass parameter, and is always less than that of empty dS
spacetime. Physically it would suggest that a de Sitter spacetime with a
black hole is more ``ordered'' than pure de Sitter spacetime, suggesting
that production of black holes in the early universe is entropically
disfavored. \ This is consistent with results on pair-production of black
holes in de Sitter spacetime \cite{MannRoss}.

\section{Conclusions}

The Gibbs-Duhem relation (\ref{entropy}) suggests that gravitational heat
should exist whenever the classical action differs from total energy.
Gravitational entropy is then given by the difference of these quantities
multiplied by the inverse temperature, provided each can be meaningfully
defined . This idea, while dating back more than 25 years to the work of
Gibbons and Hawking \cite{GibH}, is now finding application in a wide
variety of unexpected areas, made feasible because of the inclusion of the
boundary counterterm action into the quasilocal approach. These settings
include the outside of cosmological horizons and spacetimes with nonzero NUT
charge, as well as constant curvature black holes \cite{CCBH}, \ advancing
the concept of gravitational entropy into territory well beyond that of
Bekenstein's original suggestion.

Yet some troublesome questions remain. The notion of gravitational entropy
presented here rests upon the identification of the gravitational path
integral (\ref{Zpath}) with the thermodynamic partition function (\ref{helm2}%
).\ Most physicists take this identfication for granted, despite the fact
that its foundations rely far more on analogical reasoning than on
mathematical precision.\ Furthermore, although the counterterm action is
universal for both asymptotically de Sitter and anti de Sitter spacetimes,
(so that only a finite number of coefficients are available to cancel the
divergences of a given spacetime), it is not clear these are sufficient to
appropriately cancel all possible divergences \cite{EJM,baladesit}.
Moreover, the counterterm action yields a definition of total conserved
charge that is contingent upon the specific foliation of the spacetime \cite%
{shiro}. The interpretation of the various thermodynamic quantities in these
new contexts (e.g. outside of cosmological horizons) is less than clear. An
underlying statistical mechanical interpretation of these results also
remains an open question. \ And, finally, there are no obvious experiments
one might perform in the forseeable future to test such ideas, laboratory
creation of black holes notwithstanding \cite{GiddBH}.

Three decades ago Jacob Bekenstein opened the door into an exciting new
world when he pointed out the entropy/area relationship. \ The next three
decades should prove to be equally exciting.

\bigskip

{\Large Acknowledgments}

I would like to thank A.M. Ghezelbash for a careful reading of this
manuscript. This work was supported by the Natural Sciences and Engineering
Research Council of Canada.

\end{document}